# Measurement of orbital asymmetry and strain in $Co_{90}Fe_{10}$/Ni multilayers and alloys: Origins of perpendicular anisotropy*

Justin M. Shaw, Hans T. Nembach, and T.J. Silva

*National Institute of Standards and Technology, Electromagnetics Division, Boulder, Colorado 80305, USA*

We use broadband ferromagnetic resonance spectroscopy and x-ray diffraction to investigate the fundamental origin of perpendicular anisotropy in $Co_{90}Fe_{10}$/Ni multilayers. By careful evaluation of the spectroscopic $g$-factor, we determine the orbital moment along the out-of-plane and in-plane directions. For the multilayers, we find a direct relationship between the orbital moment asymmetry and the perpendicular anisotropy, consistent with the theory of Bruno [P.Bruno, Phys. Rev. B, 39, 865 (1989)]. A systematic x-ray diffraction study revealed the presence of a trigonal strain as high as 0.7 % in some samples. However, we found no direct correlation between the strain and the anisotropy, indicating that the anisotropy is not dominated by magnetoelastic effects. In order to further study the interface structure on the anisotropy, we prepared a set of equivalent alloy samples. The strain in the alloy samples was comparable to that of the multilayer samples; however the orbital moment asymmetry in the alloy samples showed a very different trend allowing us to isolate the effect of the interfaces in the multilayers.

**INTRODUCTION**

Perpendicularly magnetized materials are directly relevant to spin-transfer-torque random access memory (STT-RAM),[1,2] spin-torque oscillators (STOs),[3] and bit-patterned media (BPM).[4-10] The integration of perpendicular materials into these technologies is necessary in order to increase the thermal stability of the bit or device as these technologies are scaled to dimensions that approach 10s of nanometers.[11] In addition to thermal stability, the use of perpendicular materials has additional benefits, which include a reduction in critical currents in STT-RAM cells.[12,13] As another example, STOs using one or more perpendicular materials can generate a relatively high frequency output with little to no bias field.[14]

While perpendicular anisotropy in thin films can arise from magnetocrystalline anisotropy of highly c-axis orientated hexagonal close packed (hcp) materials[15] and atomic level superlattice materials such as the $L1_0$ order FePt and FePd,[16-20] artificially structured superlattice/multilayer structures are a convenient class of materials for many applications as a result of the tunability of both the anisotropy and saturation magnetization. These multilayers are formed by alternating 2 or more ultra-thin layers of materials many times, generating a large number of interfaces. Typical examples of multilayer materials with strong perpendicular anisotropy include Co/Pd,[21] CoFe/Pd,[22] Co/Pt,[23-26] CoNi/Pt,[27] which take advantage of perpendicular interface anisotropy caused by the electron hybridization with Pd and Pt.[28] Even though large perpendicular anisotropy is easily achieved in these multilayer materials, the presence of the Pt and Pd in the structure increases the damping parameter of the material[29-31] which can be problematic for many applications.[13] However, Daalderop *et al.* first predicted and demonstrated that a perpendicular anisotropy can also be achieved in Co/Ni multilayers.[32] This discovery was significant because all the constituents of the multilayer material are ferromagnetic 3-d transition elements. More importantly, Co/Ni was shown to have a reduced value of the damping parameter relative to many perpendicular materials that contain Pt and Pd.[33–36] In fact, we have recently shown that the damping parameter can be tuned by simply varying the ratio of Co to Ni in the multilayer.[34]

Néel first proposed the idea that an interface or surface will generate a perpendicular anisotropy due to the broken symmetry of a surface or interface.[37] However, the fundamental origin of such interface anisotropy has been the focus of much debate over the past two decades. More recently, several theoretical treatments have been directed toward understanding the microscopic origin of anisotropy in Co/Ni multilayers specifically. Daalderop *et al.* stressed the role of $d_{xy}$ and $d_{x^2-y^2}$ orbital bands in the perpendicular anisotropy of Co/Ni, which in turn also favors the [111] orientation.[32] Indeed, the perpendicular anisotropy was found to be greatly enhanced in Co/Ni and CoFe/Ni multilayers when the crystalline geometry or texture is orientated in the [111] direction.[34,38] Similarly, Gimbert *et al.* also emphasized the importance of the electronic band structure of the interface as the origin.[39] However, another source of perpendicular anisotropy may reside in magnetoelastic contributions that result from strain and lattice mismatch of the multilayer constituents.[40–43]

Many investigations on the origin of surface anisotropy have focused on the measurement of the orbital and spin moments of magnetic interfaces and surfaces.[28,44–48] The broken symmetry of an interface has long been known to enhance the orbital and spin moments at the interface relative to the "bulk" region. As an example, the calculated density of states (DOS) for a monolayer (ML) of Co on Cu differs considerably from that of bulk Co.[44] This change of the electronic structure at the interface results in an enhanced orbital moment on Co, which was measured experimentally in the same study. Recent calculations and experiments performed on Co/Ni multilayers and interfaces show an enhanced orbital moment at the Co/Ni interface.[49] However, the theoretical description reported by Bruno shows that it is the *asymmetry* in the orbital moment that gives rise to magnetocrystalline anisotropy via the spin orbit interaction.[50] Using perturbation theory applied to the tight binding model, Bruno showed that the magnetocrystalline anisotropy energy (MAE) originates from the anisotropy of the orbital moment, and becomes directly proportional to the asymmetry in the orbital moment between the hard and easy axes, as shown in Eq. (1),

$$MAE = -\alpha \frac{\xi}{4\mu_B} \Delta\mu_L \qquad (1)$$

where $\xi$ is the spin-orbit coupling parameter, $\alpha$ is the prefactor that varies between 0 and 0.2 and is a function of the electronic structure, $\mu_B$ is the Bohr magneton, and $\Delta\mu_L$ is the difference in orbital moment between the easy and hard magnetization axes. Despite the fact that this relationship is strictly valid only at 0 K, it was also shown to hold true at room temperature for some systems.[51] This treatment predicts that the easy axis will have an enhanced orbital moment as was first experimentally observed in ultra-thin Co layers[51] and later in Fe/V mulitlayers[52] and Ni/Pt multilayers.[53] However, other works do not necessarily show a direct proportionality between the MAE and orbital moment asymmetry.[41,43,54,55] This was attributed to the strong spin-orbit coupling at Au atomic sites in one case.[55] In another case this was attributed to magnetoelastic contributions and a decrease of the relative exchange splitting that occurs during the face-centered cubic (fcc) to body-centered cubic (bcc) transition.[41] Both effects can introduce an orbital moment asymmetry that deviate from Eq. (1).

Measurement of the orbital moment in materials along specific directions remains an experimental challenge. X-ray magnetic circular dichroism (XMCD) measurement performed at synchrotron facilities have long been a favored technique for the evaluation of the orbital moment. The evaluation of orbital moment asymmetry in materials with significant perpendicular anisotropy requires end chambers capable of applying large magnetic fields in multiple directions. By use of the sum rules and knowing the number of d- band holes, both the spin and orbital moments of atomic species can be determined with XMCD.[56,57] However, the orbital moment can also be determined in the laboratory using FMR through careful evaluation of the spectroscopic splitting g-factor.[52,58] The g-factor is related to the ratio of the orbital moment $\mu_L$ and the spin moment $\mu_S$ by Eq. (2)[59]

$$\frac{\mu_L}{\mu_s} = \frac{g-2}{2}.$$

(2)

FMR measurements alone cannot uniquely determine $\mu_L$ and $\mu_S$. However, if the total moment $\mu$ is known (through precise magnetometry, for example), then the separation of $\mu_L$ and $\mu_S$ becomes trivial since $\mu = \mu_L + \mu_S$. Since in the saturated state, the spin moment in this system can be assumed to be isotropic,[39] Eq. (2) shows that the orbital moment asymmetry is directly proportional to the asymmetry in the $g$–factor. For the case of perpendicular anisotropy, this orbital asymmetry is given in Eq. (3),

$$(\mu_L^\perp - \mu_L^\parallel) = \frac{\mu_S}{2}(g^\perp - g^\parallel).$$

(3)

Where, $\mu_L^\perp$ and $\mu_L^\parallel$ are the out-of-plane and in-plane orbital moments respectively. $g^\perp$ and $g^\parallel$ are the out-of-plane and in-plane $g$–factors, respectively.

Despite this relationship, determination of the orbital moment via FMR has proven to be problematic due to the difficulty of measuring the spectroscopic $g$-factor with an error less than 1 %.[58] Such an error can easily be outside the expected change in the $g$–factor resulting from a change in the orbital moment that is related to the magnetic anisotropy. However, precise measurements of the $g$-factor are made more accessible due to the increased availability of broadband network analyzers and therefore broadband FMR measurements in the laboratory.[60–62]

Using such a broadband VNA-FMR technique, we previously reported a shift in the out-of-plane spectroscopic *g*-factor in CoFe/Ni multilayers and alloys; showing a correlation of the *g*-factor with the anisotropy of the material.[34] However, we did not directly determine the relationship between anisotropy and the *g* –factor since that was beyond the scope of that particular study. Another earlier work also reported a shift in the *g*-factor for Co/Ni using an FMR technique.[33] These reports demonstrated that expected shifts in the *g*-factor due to an enhancement of the orbital moment were well within the measurement resolution of the VNA-FMR system, and therefore motivated the present study.

In this paper, we report on the precise measurement of the in-plane and out-of-plane *g*-factors in order to clarify the relationship between the perpendicular anisotropy and the asymmetry in the orbital moment. From this relationship, we test and confirm the theoretical prediction presented by Bruno.[50] To address the role of magnetoelastic effects (which can also affect the electronic structure), we perform a systematic x-ray diffraction study to determine the strain of the system and relate that to the anisotropy. Comparison of multilayer and equivalent alloy samples is used to distinguish and isolate properties that arise from the presence of the CoFe/Ni interfaces in the material.

**EXPERIMENT**

*Sample Fabrication*

Samples were dc magnetron sputter deposited at an Ar pressure of ≈ 0.5 mTorr (0.07 Pa) and a chamber base pressure of ≈ 2 × $10^{-9}$ Torr (3 × $10^{-7}$ Pa) while being rotated at approximately 1 to 2 Hz. All deposition rates were calibrated using x-ray reflectivity (XRR). All samples had a Ta (3 nm) / Cu (5 nm) seed layer to insure a strong (111) orientated crystalline texture and a Cu (3 nm) / Ta (3 nm) capping layer to prevent oxidation of the magnetic layers. The $Co_{90}Fe_{10}$/Ni multilayers consisted of 8 bi-layers of [$Co_{90}Fe_{10}(t_{CoFe})$ / Ni($r·t_{CoFe}$)] with an additional $Co_{90}Fe_{10}$ layer at the top interface. The total thickness of the multilayer is therefore $t = 8(1+r)t_{CoFe}+t_{CoFe}$. As we will see, the total multilayer thickness is a useful quantity to use when comparing properties of multilayers to those of alloys. The $(Co_{90}Fe_{10})Ni_r$ alloys were co-sputtered to produce a series of samples with identical amounts of $Co_{90}Fe_{10}$ and Ni as the multilayers over a comparable thickness range to allow for direct comparison. A schematic of the sample structures is given in Fig. 1(a). For this study, we focus on a series of multilayer and alloy samples where the CoFe to Ni thickness/composition ratio $r$ is fixed at $r = 3$ and the thickness is varied. (It is important to point out that $M_s$ will be constant for both the multilayers and alloys when $r$ is a fixed value) The use of $Co_{90}Fe_{10}$ (hereafter referred to as CoFe) instead of pure Co was used to help suppress the fcc to hexagonal close packed (hcp) transition in thicker samples, simplifying our analysis. The use of $Co_{90}Fe_{10}$ instead of Co has only a minimal effect on the anisotropy of CoFe/Ni and CoFe/Pd multilayers.

*Ferromagnetic Resonance (FMR)*

FMR measurements were performed on a 3-axis room-temperature bore superconducting magnet capable of magnetic fields as high as 3 T. The field homogeneity over the sample volume is better than 0.1 % and the field was measured at every field value using a Hall probe gaussmeter. The samples were placed face down on a co-planar waveguide (CPW) with a 50 μm center conductor. Microwave fields were generated by connecting one end of the CPW to the output port of a vector network analyzer (VNA) that had a bandwidth of 1–70 GHz. Since losses in our microwave circuit increase with frequency (necessitating increased number of averages), measurements above 50 GHz were performed only when higher frequencies were required to sufficiently increase the precision of the fitted data. The input port of the VNA was connected to the other end of the CPW and the transmission parameter (S21) through the CPW was measured at a single frequency as the external magnetic field was swept. The resonance is described by the complex susceptibility $\chi(H_{res})$ derived from the Landau-Lifshitz equation, an example of which is given in Eq. (4) for the out-of-plane geometry,[63]

$$\chi(H_{res}) = \frac{M_{eff}(H_{res} - M_{eff})}{(H_{res} - M_{eff})^2 - H_{eff}^2 - i\Delta H(H_{res} - M_{eff})},\tag{4}$$

where $H_{res}$ is the resonance field, $M_{eff}$ is the effective magnetization, $\Delta H$ is the linewidth, $H_{eff} = 2\pi f/(\gamma\mu_0)$, $f$ is the frequency, $\gamma = (g\mu_B)/\hbar$ is the gyromagnetic ratio, and $\hbar$ is the reduced Planck's constant. Of importance is the fact that the VNA-FMR technique is sensitive to both the amplitude and the phase, and therefore both the real and imaginary components to the susceptibility can be measured. Figure 1(b) shows a representative example of the spectra obtained in this study. We simultaneously fit the real and imaginary spectra to Eqn. (4) in order to determine $H_{res}$ and $\Delta H$, which is more thoroughly described in Ref. [61]. The fits of Eq. (4) to the data are also included in Fig. 1(b).

Since the samples are polycrystalline and rotated during deposition, no significant in-plane anisotropy is present in the samples and can therefore be neglected in the analysis. This assumption was verified by angular dependent in-plane magnetometry measurements performed on similar samples. We use the definition of the anisotropy energy density $E$ given by Ref. [64], which includes the second ($K_2$) and fourth ($K_4$) order perpendicular anisotropies. (The label of these anisotropy constants are inconsistent in the literature, and are sometimes labeled the first $K_1$ and second $K_2$ order anisotropy constants, respectively.)

$$E = \frac{\mu_0 M_s^2}{2}\cos^2\theta - K_2\cos^2\theta - \frac{K_4}{2}\cos^4\theta \tag{5}$$

Where, $\theta$ is the polar angle relative to the sample normal (perpendicular) direction. The Kittel equations for the perpendicular ($\perp$) and in-plane ($||$) geometries in a saturated state are given by Eqs. (6) and (7), respectively:

$$f(H_{res}) = \frac{g^{\perp}\mu_0\mu_B}{2\pi\hbar}\left(H_{res} - M_{eff}^{\perp}\right) \tag{6}$$

$$f(H_{res}) = \frac{g^{||}\mu_0\mu_B}{2\pi\hbar}\sqrt{H_{res}\left(H_{res} + M_{eff}^{||}\right)} \tag{7}$$

Using Eq. (5), the perpendicular ($M_{eff}^{\perp}$) and in-plane ($M_{eff}^{||}$) effective magnetizations become:

$$M_{eff}^{\perp} = M_s - \frac{2K}{\mu_0 M_s} = M_s - \frac{2K_2}{\mu_0 M_s} - \frac{2K_4}{\mu_0 M_s} \tag{8}$$

$$M_{eff}^{||} = M_s - \frac{2K_2}{\mu_0 M_s} \tag{9}$$

It is convenient to define the *total* perpendicular anisotropy constant $K = (K_2 + K_4)$, which is alternatively defined in Eq. (8). We use a sign convention whereby a positive value of the anisotropy constant favors a perpendicular magnetization. (Since this sign convention varies in the literature, we will adjust the sign convention of other works for consistency in our discussion.) Separation of $K_2$ and $K_4$ is straightforwardly achieved by measurement of both $M_{eff}^{\perp}$ and $M_{eff}^{||}$ and applying Eqs. (8) and (9). It is also important to point out that $M_{eff}^{\perp}$ is a measure of the net perpendicular anisotropy. For negative values of $M_{eff}^{\perp}$, the perpendicular anisotropy energy is greater than the demagnetization energy, and therefore will have a remanent magnetization that lies out-of-plane. Likewise, a sample with a positive value of $M_{eff}^{\perp}$ will have a remanent magnetization that lies in-plane.

Examples of perpendicular and in-plane geometry data with fits to the Kittel equations are given in Figs. 1(c) and 1(d), respectively. As stated earlier, precise (< 1 %) determination of the *g*-factor of thin films has remained a challenge in FMR.[58] This is particularly challenging for the in-plane geometry since, as Eq. (7) shows, there is a non-linear relationship between *f* and $H_{res}$ unless $H_{res} \gg M_{eff}^{//}$. As a result, measurements must be performed over a large range of *f* and $H_{res}$ in order to obtain reliable fits. To overcome these challenges, we apply asymptotic analysis using the methods outline in Ref. [[62]] in order to increase the precision of our measurements of the *g*-factor.

*X-ray diffraction (XRD)*

In-plane and out-of-plane XRD measurements were performed on a diffractometer equipped with a 4-circle goniometer and an instrument resolution of 0.0001°. In both cases, we used a Cu $K_\alpha$ source with parallel beam optics. A powder Si samples was used to calibrate the 2θ angle and adjust for any offsets. The out-of-plane lattice constant was determined from fits to the (111) fcc peak, as shown in Fig 2(a). Analyses of the (111) peaks are complicated due to the presence of intense thickness fringes; predominately on the lower angle side of the (111) peak. As a result, multiple pseudo-Voigt functions were required to fit the data, which are included in Fig 2(a). Of importance is the fact that (excluding the thickness fringes) there is no evidence of a shoulder or multiple peaks in the spectra and therefore conclude that the system is predominately uniformly strained. This observation is consistent with measurements in epitaxial Co/Ni,[49] but is in contrast to another recent work on sputtered Co/Ni, which shows relaxation between the various layers within the multilayer.[33] However, in the latter work, a Pt layer was used in the seed layer structure, which introduces significantly more strain (> 10%) into the system since Pt is poorly lattice matched with Cu, Co, or Ni. One might argue that the peak on the higher angle side is not a thickness fringe, but rather a shoulder that results from relaxation of the material. However, we conclude that this higher angle peak is indeed a thickness fringe for 2 reasons: (1) the peak position of the higher angle peak is shifted away from the expected peak locations of Cu, CoFe, or Ni. Relaxation would be expected, by definition, to shift the lattice constant towards the unstrained bulk value; and (2) plots of the relative peak positions as a function of thickness indicate that it behaves as a thickness fringe and is even present in samples with only a single Cu layer (see Appendix I).

The in-plane lattice constant was determined by fits to the (220) fcc peak, as shown in Fig. 2(b). In contrast to the out-of-plane data, the in-plane data show the presence of two distinct peaks, which were each fit to a pseudo-Voigt function. We assign the higher angle peak to the CoFe/Ni layer and the lower angle peak to that of the Cu layer(s). This assumption is validated by the variation of the peak intensities as a function of grazing incidence angle, thereby producing a rough depth profile through the material. (see Appendix I). Further confirmation of this assumption is given by comparison of the relative peak intensities; the relative intensity of the lower angle peak decreases substantially as the thickness of the CoFe/Ni layer is increased, as would be expected if the Cu thickness is constant. (see Appendix I) These data indicate that the Cu layer is not uniformly strained with the CoFe/Ni layer in the in-plane direction, but the CoFe and Ni layers are uniformly strained within the CoFe/Ni layers.

Texture analysis confirmed that all samples in this study were highly (111) textured with rocking curve widths of 2.4 – 3.6 ° full width at half maximum. Cross-section transmission electron microscopy images on similar samples have indicated that the Ta layer is amorphous and that the Cu-CoFe-Ni system consists of quasi-coherent, [111]-orientated columnar grains. The amorphous structure of the Ta is also validated by the absence of Ta diffraction peaks in the XRD analysis.

**RESULTS**

## A. Anisotropy

Figure 3(a) shows a plot of the perpendicular $M^\perp_{eff}$ as a function of the reciprocal thickness for both the multilayer and alloy samples. Recall that negative values for $M^\perp_{eff}$ indicate that the material has a net perpendicular anisotropy, and therefore the remanent magnetization will be out-of-plane. As expected for the multilayers, the perpendicular anisotropy is sufficiently large in the samples where t < 10.4 nm ($t_{CoFe}$ < 0.3 nm) to have a net perpendicular anisotropy. In fact, $M^\perp_{eff} = 0$ for the sample with t = 10.4 nm ($t_{CoFe}$ = 0.3 nm) (i.e. the demagnetization energy is equal to the perpendicular anisotropy energy and $M^\perp_{eff} = 0$). For the thicker multilayer samples, the remanent magnetization lies in-plane. Unlike the multilayers, the net anisotropy of all the alloy samples lies in-plane.

Figures 3(b) and 3(c) are plots of $K_2$ and $K_4$ as a function of the reciprocal thickness $1/t$ of the multilayer and alloy samples. The strong $1/t$ dependence of the anisotropy in the multilayer samples suggests that the perpendicular anisotropy is largely interfacial in origin (or at least originates from another material property that also has a $1/t$ dependence). These data also show that $K_4$ is significantly smaller than $K_2$, which is consistent with previous reports.[33,65]

With the assumption of an interface anisotropy, we can straightforwardly separate the volume (or bulk) anisotropy component from the interface component of $K_2$ and $K_4$ in the multilayer data using the phenomenological equation given in Eq.(10).

$$K(t) = K^{vol} + \frac{(2n+1)K^{int}}{t},$$

(10)

where $K^{vol}$ is the volume anisotropy term, $K^{int}$ is the interface anisotropy term, and $n = 8$ is the number of bi-layers in the multilayer. We note that the interface anisotropy determined from this equation is an *average* over all the interfaces, and it is likely that the CoFe/Cu and CoFe/Ni interfaces have different values. We apply Eq. (10) to both $K_2$ and $K_4$ data to obtain the respective volume and interface components, which are listed in Table I.

The value of $K_2^{int}$ agrees well with the value of $3.1 \times 10^{-4}$ J/m$^2$ reported by Daalderop *et al.*[32] in evaporated Co/Ni multilayers. The slightly lower value of $K_2^{int}$ that we measure may be a result for several factors (and in general will affect direct comparison between any two studies). First, we use Co$_{90}$Fe$_{10}$ instead of pure Co, which may alter the electronic structure. Second, Daalderop *et al.* used a larger number of bi-layer repeats; decreasing the influence of the Co/Cu interface. Thirdly, sputter deposition used in our study (versus evaporation) may cause increased intermixing of the interface, decreasing the anisotropy. The effect of intermixing was explored in ion irradiated Co/Ni multilayers that displayed a decreased perpendicular anisotropy as intermixing of the interface was intentionally induced by the ions.[66] In addition, the growth parameters, such as the seed layer, has been shown to have an influence on the anisotropy of Co/Ni.[67,68] Slightly larger anisotropy energy for example, was also reported in single crystal epitaxially grown Co/Ni multilayers[69] Daalderop *et al.* deposit the Co/Ni on thick Au underlayers, whereas we use a thin Ta/Cu bilayer as a seed layer. Finally, sputter deposition may produce a different density of stacking faults in the material versus evaporation. Stacking faults were shown to alter the anisotropy in calculations performed on Co/Ni multilayers.[39]

**Table I. The interface and volume components to $K_2$ and $K_4$ for the multilayer samples.**

|  | Volume component (J/m$^3$) | Interface component (J/m$^2$) |
|---|---|---|
| $K_2$ | $K_2^{vol} = -(0.76 \pm 0.17) \times 10^5$ | $K_2^{int} = (2.83 \pm 0.02) \times 10^{-4}$ |
| $K_4$ | $K_4^{vol} = (0.34 \pm 0.03) \times 10^5$ | $K_4^{int} = -(0.24 \pm 0.03) \times 10^{-4}$ |

Also plotted in Figs. 3(b) and 3(c) are the values of $K_2$ and $K_4$ as a function of $1/t$ for the alloy samples. As expected, the perpendicular anisotropy of the alloy samples is significantly smaller than the multilayers samples. However, the $1/t$ dependence still indicates the presence of an interface anisotropy. This is not surprising considering that the alloy samples have 2 interfaces with Cu; one with the capping layer and one with the seed layer of the material. We also determine the components to the anisotropy using the following equation: $K(t) = K^{vol} + 2K^{int}/t$. This equation takes into account that there are only 2 interfaces in the alloy samples. While no previous reports have been conducted on the interface anisotropy of the CoFeNi/Cu or CoFe/Cu interface, the Co/Cu(111) interface was reported to have an interface component of $\approx 1 \times 10^{-4}$ to $2 \times 10^{-4}$ J/m², which is consistent with the value of $(1.56 \pm 0.03) \times 10^{-4}$ J/m² that we measure for the CoFeNi/Cu interface.[70-73] This also indicates that the CoFe/Cu interface anisotropy is approximately half that of the Co/Ni interface in the multilayers.

**Table II. The interface and volume components to $K_2$ and $K_4$ for the alloy samples.**

| | Volume component (J/m³) | Interface component (J/m²) |
|---|---|---|
| $K_2$ | $K_2^{vol} = (0.03 \pm 0.01) \times 10^5$ | $K_2^{int} = (1.56 \pm 0.03) \times 10^{-4}$ |
| $K_4$ | $K_4^{vol} = (0.03 \pm 0.01) \times 10^5$ | $K_4^{int} = -(0.19 \pm 0.02) \times 10^{-4}$ |

### B. Strain

The in- and out-of-plane lattice constants for the magnetic layer are plotted in Fig. 4(a) as a function of the reciprocal thickness $1/t$. Also included in the plot (red line) is the calculated lattice constant obtained for an ideal, fully strained system of Cu, CoFe, and Ni, using Vegard's law[74] and the bulk values of the respective lattice constants. We do not include Ta in this calculation since our previous XRD and TEM analysis have indicated that Ta is an amorphous phase and is therefore not part of the columnar grains that are formed in the Cu/CoFe/Ni system. We also indicate the location of the lattice constants of bulk Cu and the Co-Fe-Ni (also calculated using Vegard's law) for reference. Both the in-plane and perpendicular lattice constants of the CoFe/Ni layer show a partial relaxation, but approach that of the ideal value for the CoFeNi system as the thickness increases. This behavior is expected since the strain induced by the Cu layers will diminish as the magnetic layer thickness becomes large relative to the Cu thickness. For the thinnest samples, the lattice constants begin to approach the Cu-CoFe-Ni curve, but always remain partially relaxed. Of importance is the fact that the effective in-plane and out-of-plane lattice constants for all but the thickest samples are different. This behavior indicates the presence of a significant trigonal distortion in the magnetic layer that could give rise to a magnetic anisotropy via magnetoelasticity.

We therefore calculate the anisotropic strain in the system as the difference between the perpendicular and in-plane lattice constants, which is plotted in Fig. 4(b). Immediately, we see that the strain in the multilayer system does not obey a strict $1/t$ dependence, and in fact, begins to saturate at approximately $1/t \approx 0.13$ nm$^{-1}$ ($t_{CoFe} \approx 0.23$ nm). We also measured the strain in the alloy samples of similar composition, which is also included in Fig. 4(b). Of importance is the fact that there is little difference between the measured strain in the multilayers and the alloys, yet the multilayers have approximately an order of magnitude or greater value of perpendicular anisotropy relative to the alloys. These data provide strong evidence that magnetoelastic effects do not play a dominant role in the perpendicular anisotropy in CoFe/Ni multilayers. This conclusion is further substantiated by another recent XRD study in Co/Ni multilayers that also found that the calculated magnetoelastic component to the anisotropy cannot account for perpendicular anisotropy.[49]

## C. Orbital moment and asymmetry

We now turn our attention to the relationship between the orbital moment and the anisotropy. Figure 5(a) shows a plot of both the in-plane and out-of-plane *g*-factors as a function of the reciprocal thickness $1/t$ for the multilayer samples. For the out-of-plane data, the *g*-factor increases slightly as the thickness of the multilayer decreases (or alternatively, the reciprocal thickness increases). From Eq. (3), we see that this behavior translates into an increase in the perpendicular orbital moment of the material as the thickness decreases. In contrast, the in-plane *g*-factor decreases as the thickness of the multilayer decreases. It is interesting to point out that the change in the in-plane *g*–factor is approximately 4 times greater relative to the out-of-plane direction. In both cases, the *g*–factor exhibits a strongly correlated dependence to the reciprocal thickness, consistent with the effect originating at the interface. Furthermore, this behavior suggests that the geometric confinement causes a perturbation of the electron orbits at the interface.

The values of the *g*-factor for the alloy samples, plotted as a function of the reciprocal thickness, are given in Fig. 5(b). An asymmetry in the orbital moment is also observed for the alloy samples, however the opposite trend is observed: the out-of-plane and in-plane *g*-factors decreases and increase respectively, with decreasing thickness. However, the change in the *g*-factor for the in-plane case is still greater in magnitude than that for the out-of-plane case. Recall that the alloy samples have two interfaces and therefore some interface effects may still be present.

To more directly examine the relationship between the orbital asymmetry and the anisotropy, we first rewrite Eq. (1) into a form that is more convenient for our particular experiment and set of parameters, which is given in Eq.(11).

$$K = -\alpha \frac{\xi N}{8V} \frac{\langle \mu_S \rangle}{\mu_B} \left( g^\perp - g^\| \right) = -\left( \alpha \frac{\xi N}{4V} \right) \frac{\Delta \mu_L}{\mu_B} \; , \tag{11}$$

where $K$ is the perpendicular anisotropy energy density in J/m³, $V$ is the volume of the unit cell, $N$ is the number of atoms per unit cell, $\Delta\mu_L = (\langle\mu_L^\perp\rangle - \langle\mu_L^\parallel\rangle)$, $\langle\mu_L^\perp\rangle$ is the average orbital moment in the perpendicular direction, and $\langle\mu_L^\parallel\rangle$ is the average in-plane orbital moment. With the assumption that $\langle\mu_S\rangle$ shows little variation as a function of thickness, this equation indicates that $K$ and $\Delta\mu_L$ should be proportional. We validate the assumption that $\langle\mu_S\rangle$ shows little variation as a function of thickness through magnetization measurements performed with a magnetometer based on a superconducting quantum interference device (SQUID), which show no variation of $M_s$ with the thickness of the multilayers (see Appendix II). In addition, recent XMCD measurements taken on sputtered Co/Ni multilayers also did not indicate a significant change of $\mu_S$ with thickness, despite that fact that in that same study an enhanced spin moment was observed at the interface in epitaxial Ni/Co/Ni trilayers.[49] From our SQUID measurements, we estimate $\langle\mu_S\rangle$ to be 0.83±0.02 $\mu_B$.

We plot the orbital moment asymmetry $\Delta\mu_L/\mu_B$ as a function of $K$ for the multilayer samples in Fig 6(a). For the multilayers, the relationship between the orbital moment asymmetry and the anisotropy is in agreement with the tight binding perturbation theory established by Bruno et al..[50] From the slope of this curve we calculate the prefactor to Eq. (11) as $\alpha$ = 0.097 ± 0.007. We use the value of $\xi$ = −1.58 × 10⁻²⁰ J/atom (0.10 eV/atom) for the spin-orbit coupling parameter, which is calculated as the weighted average of the spin-orbit coupling parameters of Ni, Fe, and Co in the material.[75] To put this value of $\alpha$ in perspective, previous measurements of $\alpha$ were determined to be 0.2 for ultrathin Au/Co/Au trilayers[51], 0.1 for Ni/Pt multilayers[53], and 0.05 for Fe/V multilayers.[52] The value we measure for the CoFe/Ni system is therefore within the range of values measured in other material systems, and that predicted by theory.

This model, however, breaks down for the alloy system. Figure 6(b) shows a plot of the orbital asymmetry as a function of $K$ for the alloy samples. Unlike the multilayers, these data exhibit a negative slope and a non-zero $y$-intercept. Regardless, if we calculate a prefactor to Eq.(11) solely from the slope of the data, it becomes $\alpha = -0.020 \pm 0.005$. In addition to being negative, the magnitude of $\alpha$ is approximately a factor of 5 smaller than that of the multilayers.

**DISCUSSION**

Eq. (1) predicts that the orbital moment asymmetry is proportional to the magnetic anisotropy, and that the direction of enhanced orbital moment is directed along the easy axis.[50] In addition, recent first-principle calculations also confirmed that such an asymmetry in the orbital moment should be present in the Co/Ni multilayer where the enhancement is along the perpendicular direction.[39] We have therefore been able to confirm both of these predictions; showing that an orbital moment enhancement is present in the perpendicular direction and that the orbital asymmetry is proportional to the perpendicular anisotropy. More importantly, this proportionality was confirmed for the first time over a tenfold range of anisotropy.

It is important to point out that the relationship between $K$ and $\Delta\mu_L$ in Eq. (11) is strictly valid only when $K$ is measured at 0 K. However, we surmise that Eq. (11) still holds true in our case due to the fact that the Curie temperature $T_c$ is much higher than room temperature (RT). While we were unable to perform SQUID measurements to high enough temperature to directly measure $T_c$, measurements taken to 400 K show only a 7 % to 10 % reduction in $M_s$ relative to the value of $M_s$ taken at 10 K. This indicates that $T_c$ is much higher than 400 K and the reduced temperature $T/T_c$ is approximately constant for the samples in this study. In addition, there is no significant variation in this value among either the multilayer or alloy samples in this study. As a result, we speculate that $K$ taken at room temperature (298 K) is approximately reduced by a similar factor for all samples in the study. (i.e. $K_{0K} \propto K_{300K}$). Under this assumption, Eq. (11) will still hold at RT, but the value of the prefactor $\alpha$ at RT may be reduced.

The data for the alloy samples differs from the multilayer data in that the perpendicular direction is not always the direction of the orbital moment enhancement. Since the alloy samples have two interfaces with Cu, these data suggest that the CoFeNi/Cu interfaces provide a different perturbation to the orbital motion relative to the CoFe/Ni interface as similarly observed in Au/Co/Au trilayers.[55] Alternatively, magnetoelastic effects were shown to generate an asymmetry in the orbital moment that is not necessarily linear with anisotropy.[41] However, by comparison between the multilayer data and the alloy data, we can conclude that the presence of the CoFe/Ni interfaces cause a significant asymmetry in the orbital moment and perpendicular anisotropy that are related through Eq. (1).

As a final point of discussion, we address the fact that first-principle calculations have predicted that the anisotropy energy will vary greatly depending on the ratio of Co to Ni. This effect was explained to result from either the location of the Fermi energy in the band structure[32] or peaks in the density of states that favor certain virtual transition that can occur between electron orbitals near the Fermi energy.[39] For example, such calculations predict a strong perpendicular anisotropy for a multilayer consisting of 1 monolayer (ML) Co and 2 ML Ni ($r = 2$), but a much reduced perpendicular anisotropy for a multilayer consisting of 1 ML Co and 5 ML Ni ($r = 5$).[32] We therefore deposited two additional thickness series of samples with $r=2$ and $r=5$ in order to explore this prediction. The anisotropy constants of the samples along with the $r=3$ samples used in this study are plotted versus the reciprocal thickness in Fig. 7. Surprisingly, there is no discernible difference in the data between samples with different values of $r$, within the scatter of the data. As a result, these data show that differing values of CoFe and Ni content have minimal effect on the interface anisotropy; simplifying the picture of the perpendicular interface anisotropy (at least within the range of $r = 2$ to $r = 5$). Thus, within this range of values of $r$, we find that the perpendicular anisotropy is determined solely by the interface density of the material.

**SUMMARY**

We have reported on a systematic FMR and XRD study in CoFe/Ni multilayers and alloys. An equivalent trigonal distortion was measured in both the multilayers and alloys despite very large differences in perpendicular anisotropy between the two sample structures. This fact, coupled with the lack of correlation between the strain and anisotropy strongly suggests that the strain is not the primary origin of perpendicular anisotropy. The orbital moment asymmetry of the samples was measured via precise determination of the *g*-factor in both the in-plane and out-of-plane directions. For the multilayers, we found a direct proportionality between the perpendicular anisotropy and the orbital moment asymmetry over a range that exceeds an order of magnitude in the anisotropy. This proportionality is consistent with the theoretical predictions of Bruno.[50] The alloy samples also show a linear relationship between the orbital moment and anisotropy; however, the magnitude of this relationship is a factor of 5 lower and of opposite sign relative to the multilayers. We surmise that the trend in the orbital asymmetry in the alloys originates in the strain and/or the interface with the seed and capping layers. By comparison of the multilayer and alloy results, we conclude that the presence of the CoFe/Ni interface produces an asymmetry in the orbital moment due to the electronic structure localized at the interface, and that this asymmetry is the primary origin of perpendicular anisotropy in the system.

**ACKNOWLEDGMENTS**

The authors are grateful to Tony Kos for help with the instrument setup, and Michael Schneider, Carl Boone, Mathias Weiler and Phil Ryan for valuable discussions.

**APPENDIX I: X-ray diffraction**

Figure 8(a) shows a plot of the (111) peak for a series of multilayer samples. The presence of the high intensity thickness fringes are seen on the low angle side of the primary peak (peak of highest intensity). As expected, as the thickness of the multilayer increases, the period of the fringes decrease until at $t_{CoFe}$ = 0.75, the fringes can no longer be resolved due to the period becoming smaller than the linewidth of the peaks. The thickness fringes on the higher angle side of the primary peak are much lower in intensity, and therefore are more difficult to resolve. Also present in the spectra are the superlattice peaks, which are labeled in the plots as ±1, ±2, and ±3 for the first, second, and third order peaks, respectively.

The thickness fringes along with the primary peak were simultaneously fit to multiple pseudo-Voigt functions as demonstrated in Fig. 2. The position of the primary peak and the thickness fringes are plotted as a function of $t_{CoFe}$ in Fig. 8(b). Here, the trend in the thickness fringes is further clarified. Of importance is the presence of both the higher angle and the lower angle fringes in the $t_{CoFe}$ = 0 (i.e. 8 nm Cu only) sample. In addition, the location of Cu, CoFe, and Ni are also indicated as the horizontal lines in the plot. The position of the thickness fringes do not correspond to the expected peak locations of a simple relaxation model. Therefore, these data unambiguously show that the origin of the peaks on both sides of the primary peak are thickness fringes and not due to multiple peaks resulting from relaxation of the multilayer system.

A similar plot of several in-plane (220) peaks for multilayer samples of various thicknesses is shown in Fig. 9(a). Two distinct peaks are observed. The intensity of the lower angle peak decrease relative to the higher angle peak as the thickness of the multilayer increases. Since the lower angle peak is also closest to the ideal peak location for Cu, we conclude that the lower angle peak comes from Cu and the higher angle peak comes from the CoFe/Ni layer. However, we cannot resolve whether the lower angle peak includes one or both the bottom and top Cu layer. We can speculate that the bottom 5 nm Cu layer forms with an in-plane lattice constant closest to the ideal value of Cu. Since the top 3 nm Cu layer is formed on the CoFe/Ni layer, it is possible that it is strained with the CoFe/Ni, and has a different lattice constant from that of the bottom Cu layer.

To attempt to resolve this question, we perform a crude depth profile of the in-plane (220) peak for the $t_{CoFe}$ = 0.2 sample. This is achieved by varying the tilt angle $\Psi$ of the x-ray beam at grazing incidence, just above the critical angle. (By our definition of $\Psi$, out-of-plane measurements correspond to $\Psi$ = 0 °, and in-plane measurements correspond to $\Psi$ = 90 ° where the x-ray beam is parallel to the surface.) The critical angle is the angle at which the x-ray will begin to penetrate the surface of the sample. As the angle is increased, the x-rays will penetrate deeper into the sample. This effect can be understood in the data shown in the inset of Fig. 9(b). Here, the intensity of the (220) peak is plotted as a function of the tilt angle. The intensity is at background level until about 90.2 °, where it rapidly increases as $\Psi$ increases. (Indicating that the critical angle is approximately 0.2 °) The intensity increases until about 90.5 ° and then begins to decrease again as the x-rays penetrate into the substrate. If we take spectra at various points between 90.2 ° and 90.5 °, then we are essentially diffracting from different ranges within the material. To demonstrate this, in Fig. 10(b) we take spectra at 3 different values of $\Psi$, which are indicated in the figure inset. The three spectra are normalized such that the high angle peaks are of equal intensity. As spectra are taken that progressively penetrate deeper into the sample.

Of importance is that the blue data is a spectrum taken from approximately the top half of the material (or slightly less), which shows the presence of the Cu in the spectrum; likely originating from the top Cu layer. As spectra are taken where the x-rays progressively probe deeper into the material, the Cu peak increases in intensity, as would be expected as more of the beam is diffracted from the bottom 5 nm Cu layer. This observation provides strong evidence that both the Cu layers are partially relaxed in the in-plane direction.

**APPENDIX II: SQUID measurements**

The temperature dependent (10-400 K) and room temperature (298 K) magnetization were performed using SQUID magnetometry. Samples were diced into 6 mm × 6 mm pieces and measurements were performed with an external field applied in the film plane. The inset in Fig. 10 shows a typical magnetization curve for the samples studied in this work. Magnetization curves were performed to fields at least 2 T higher than the saturation field to ensure the diamagnetic background from the Si substrate can be accurately subtracted. $M_s$ was calculated from the sample size and thickness of the multilayer or alloy sample. Figure 10 is a plot of $M_s$ as a function of thickness. Within the scatter of the data, the value of $M_s$ does not vary with thickness. Since $\mu_S \gg \mu_L$, we can conclude that $\mu_S$ is constant for all samples in this study. The slightly lower value of $M_s$ relative to the ideal value calculated using a weighted average of the bulk values of $M_s$ for the individual constituents is likely a result of deviations of deposition rates.

Figure 11 is a plot of $M_s$ normalized to the saturation magnetization taken at T =10 K ($M_{10K}$) for the $t_{CoFe}$ =0.4 nm multilayer sample. At 400 K the magnetization is reduced by approximately 8.8 % relative to the value at 10 K. This value varied from 7 to 10 % over several samples in this study indicating that the Curie temperature is much larger than room temperature and that the Curie temperature is similar among samples.

**FIGURE CAPTIONS**

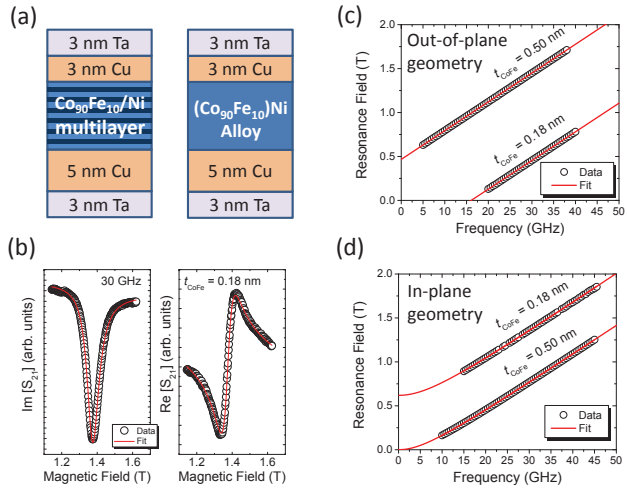

**Figure 1.** (a) Schematic diagram of the multilayer and alloy sample structures. (b) Examples of imaginary and real parts of the FMR spectra taken at 30 GHz for the $t_{CoFe}$ = 0.18 nm multilayer sample. The line through the data is the fit to Eq. (4). Example Kittel plots of the resonance field versus frequency for the (c) out-of-plane and (d) in-plane geometries. The lines through the data are fits to Eqs. (6) and (7).

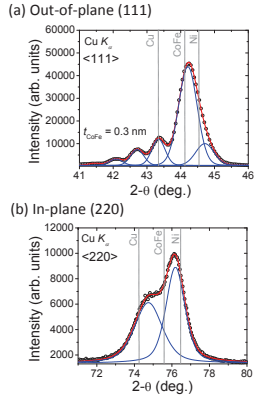

**Figure 2.** XRD spectra of the (a) out-of-plane (111) peak and (b) in-plane (220) peak taken on the $t_{CoFe}$ = 0.3 nm multilayer sample. The individual and combined pseudo-Voigt function fits are included as the solid lines. The locations of the ideal Cu, Ni, and CoFe peak positions are included as the vertical dotted lines.

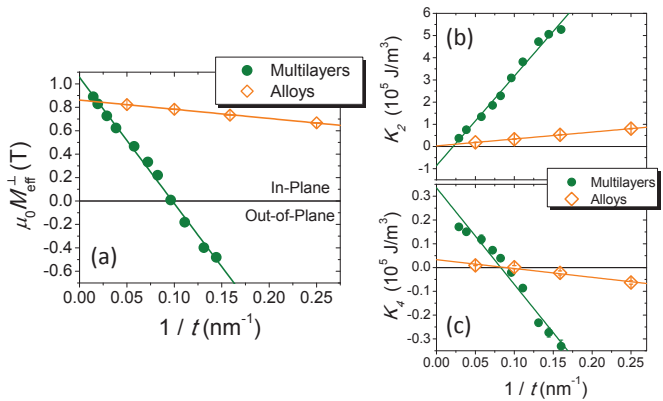

**Figure 3.** (a) Plot of the perpendicular effective magnetization as a function of reciprocal thickness for both the multilayer (solid circles) and alloy (open diamonds) samples. The (b) second order and (c) fourth order anisotropy constants are plotted as a function of reciprocal thickness. The linear fits through the data were used to separate out the respective interface and volume contributions.

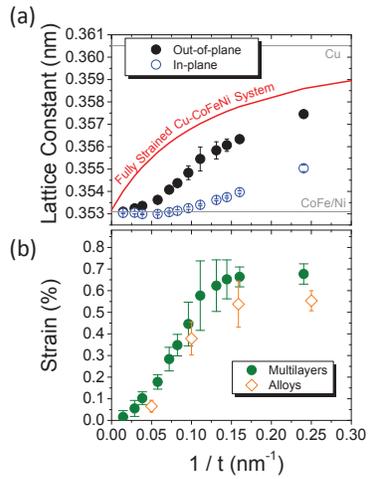

**Figure 4.** (a) The effective out-of-plane and in-plane lattice constants as a function of reciprocal thickness for the multilayer samples. The solid curve is the calculated lattice constant for a fully strained system. The lattice constants for pure Cu and CoFeNi are also indicated. (b) The measured strain as a function of reciprocal thickness for the multilayer and alloy samples.

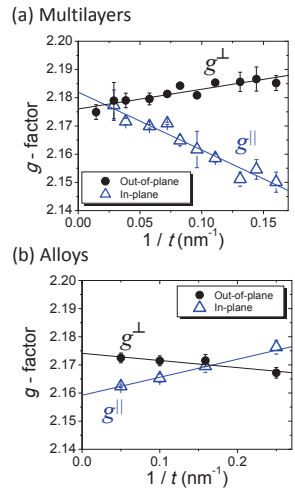

**Figure 5.** The out-of-plane (solid circles) and in-plane (open triangles) values of the *g*-factor versus the reciprocal thickness for the (a) multilayer, and (b) alloy samples.

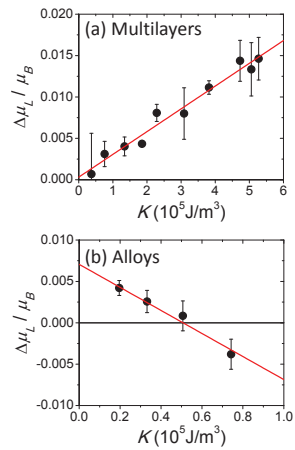

**Figure 6.** The orbital moment asymmetry versus the perpendicular anisotropy constant for the (a) multilayer, and (b) alloy samples. Linear fits through the data used to calculate the prefactor to Eq. (1) are included in the plots as the solid lines.

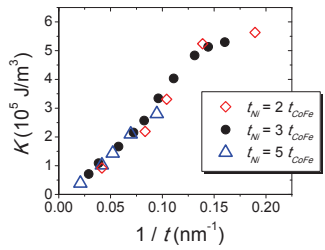

**Figure 7.** The perpendicular anisotropy constant as a function of the reciprocal thickness for multilayers with different ratios of Ni thickness to CoFe thickness.

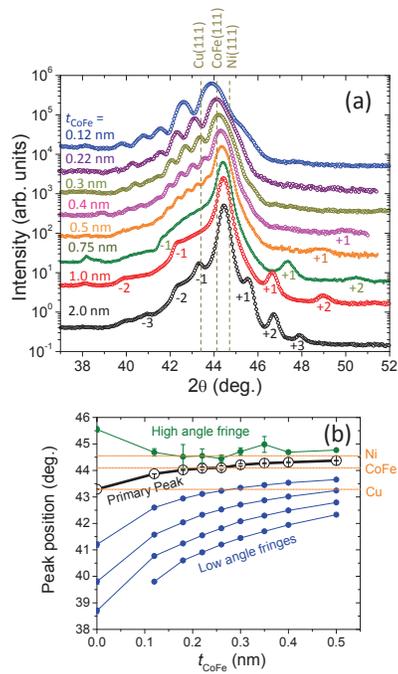

**Figure 8.** (a) Plots of the intensity on a logarithmic scale versus 2θ for the out-of-plane (111) peak taken of the multilayer samples. The presence of thickness fringes are seen predominantly on the lower angle side. Superlattice peaks are also visible in the spectra, which are labeled for the first ±1, second ±2, and third ±3 orders. (b) The position of the primary (111) peak and the thickness fringes is plotted as a function of $t_{CoFe}$. The peak locations of pure Ni, $Co_{90}Fe_{10}$, and Cu are included for reference.

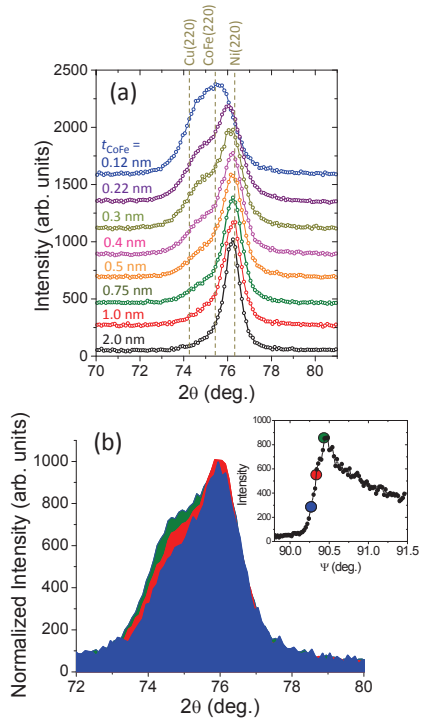

**Figure 9.** (a) Plots of the intensity versus 2θ for the in-plane (220) peak taken of the multilayer samples. (b) Plot of the in-plane (220) peak for the $t$=0.22 nm multilayer sample at various grazing incidence angles. The inset shows the intensity of the (220) peak as a function of the tilt angle $\Psi$, and indicate the points on the curve where the (220) spectra were taken.

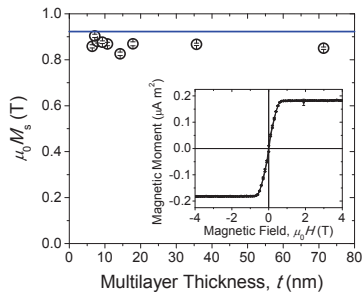

**Figure 10.** Plot of $M_s$ as a function of the multilayer thickness. The horizontal line indicates the ideal value of $M_s$ assuming a weighted average of $M_s$ between constituents in the multilayer. The inset is an example of a SQUID magnetization curve taken of the $t_{CoFe}$ = 0.18 nm multilayer sample with an in-plane external magnetic field.

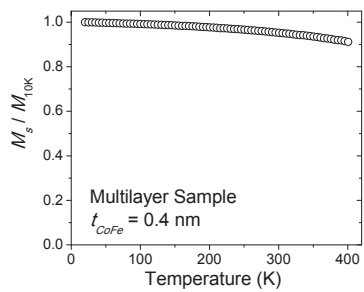

**Figure 11.** Plot of $M_s$ versus temperature for the $t_{CoFe}$=0.4 nm multilayer sample.